\documentclass[12pt]{article}\usepackage[hyperfootnotes=false]{hyperref}
\usepackage{epsfig}
\usepackage{float}
\usepackage{empheq}
\usepackage{bbold}

\usepackage[utf8]{inputenc}
\usepackage{amsmath}

\usepackage{caption}

\usepackage{amsmath}
\usepackage{amssymb}
\usepackage{graphicx}
\newlength{\abstractwidth}
\renewcommand{\thefootnote}{\fnsymbol{footnote}}
\renewcommand{\thanks}[1]{\footnote{#1}} 
\newcommand{\starttext}{
\setcounter{footnote}{0}
\renewcommand{\thefootnote}{\arabic{footnote}}}

\newcommand{\be}{\begin{equation}}
\newcommand{\bea}{\begin{eqnarray}}
\newcommand{\eea}{\end{eqnarray}}
\newcommand{\beq}{\begin{equation}}
\newcommand{\ee}{\end{equation}}

\newcommand*\widefbox[1]{\fbox{\hspace{2em}#1\hspace{2em}}}

\def\eq{&=&}

\def\simleq{\; \raise0.3ex\hbox{$<$\kern-0.75em
\raise-1.1ex\hbox{$\sim$}}\; }
\def\simgeq{\; \raise0.3ex\hbox{$>$\kern-0.75em
\raise-1.1ex\hbox{$\sim$}}\; }

\def\bi{\begin{itemize}}
\def\ei{\end{itemize}}
\def\S{Schwarzschild}

\def\dof{degrees of freedom }

\def\CJ{{\cal{J}}}

\def\t{\tau}

\def\bsub{ \begin{subequations}
\begin{empheq}[box=\widefbox]{align}  }
\def\esub{ \end{empheq}
\end{subequations}}

\def\1{\(  \mathbb{1} \)}

 \def\lf{\left(}
    \def\rg{\right)}

  \def\bn{\bigskip \noindent}

 \def\bm{\begin{bmatrix}}
 \def\em{\end{bmatrix}}

    \def\J-{\CJ^{-1}}
    
       \def\o{\omega}

\makeatletter
\g@addto@macro\normalsize{%
  \setlength\abovedisplayskip{10pt}
  \setlength\belowdisplayskip{20pt}
  \setlength\abovedisplayshortskip{10pt}
  \setlength\belowdisplayshortskip{20pt}
}
\makeatother

\usepackage{color}

\begin{document}


  
\begin{titlepage}

 \rightline{}
\bigskip
\bigskip\bigskip\bigskip\bigskip
\bigskip

\centerline{\Large \bf {Scrambling in Double-Scaled SYK and De Sitter Space     }} 

\bn

\bigskip
\begin{center}
\bf      Leonard Susskind  \rm

\bigskip
Stanford Institute for Theoretical Physics and Department of Physics, \\
Stanford University,
Stanford, CA 94305-4060, USA \\

and

Google, Mountain View, CA

\end{center}

\bn

\begin{abstract}

 I want to call attention to a simple previously noted fact about the double-scaled version of the SYK model which suggests that it may be holographically dual to de Sitter space.

\end{abstract}

\end{titlepage}

\starttext \baselineskip=17.63pt \setcounter{footnote}{0}

\Large



\section{Double-Scaled Model}
In the conventional SYK model the ``locality parameter" $q$ is held fixed as $N\to \infty.$ By contrast, the double-scaled model \cite{Cotler:2016fpe}\cite{Berkooz:2018jqr}
is defined by letting  $\frac{q^2}{N}$ tend to a constant as $N\to \infty.$

In this paper I will mean something a bit more general by the double-scaled limit. I will call a model double-scaled if $q\to \infty$ as any power of $N$, greater than $0$ and less than $1.$ This of course includes the usual double-scaled model but not the conventional fixed-$q$ model. I will also be considering the high-temperature limit in which the scrambling process may be described by an epidemic model. 

In the original  literature on the double-scaled model  \cite{Cotler:2016fpe}\cite{Berkooz:2018jqr} I was unable to find any discussion of what the bulk gravitational dual might be. But in
 a recent paper \cite{Susskind:2021esx}  I briefly conjectured that the high-temperature, double-scaled model is dual to a de Sitter space version of $JT$ gravity with a positive cosmological constant.  A similar unpublished claim was made by H. Verlinde \cite{Verlinde}.
 
 In this note I want to call attention to one piece of evidence for such a duality. It was  briefly described in \cite{Susskind:2021esx}, but may have been overlooked. 
 
 \section{Hyperfast Scrambling}
 I will use the ``epidemic" model 	\cite{Susskind:2014jwa}\cite{Roberts:2018mnp}  to describe the scrambling properties of SYK. We consider $N$ qubits where all the qubits are initially healthy. One extra ``sick" qubit is added to the system at time $\t=0.$ At time $\t=1$ the qubits randomly come together in groups of $(q-1),$ and if one in the group is sick the rest in that group become sick. In the next round the qubits are randomly regrouped into groups of $(q-1)$ and again meet at time $\t=2.$ Obviously the epidemic grows. The ``size" of the epidemic at time $\t$  is called $\sigma(\t).$  It  is defined to be the number of sick qubits at time $\t$. The fraction of the qubits which are sick is $P(\t).$ The relation between $P$ and $\sigma$ is,
 \be 
 \sigma = N P.
 \label{s=NP}
 \ee
 
 \subsection{Recursion Relation and Solution}
 
 $P(\t) $ satisfies the following recursion relation\footnote{Adam Brown, unpublished. A smoother behavior can be obtained by allowing the qubits to come together after time $\epsilon$ and infect another qubit with probability $\epsilon.$ The recursion relation in this case is $$P(\t +\epsilon) = P(\t)+ \epsilon[1-P(\t)] \ [   1-\big(  1-P(\t) \big)^{q-1}]$$ In the limit $\epsilon \to 0$ it gives the continuum limit in equation \eqref{diffeq}.}.
 \be 
P(\t +1) = P(\t)+[1-P(\t)] \ [   1-\big(  1-P(\t) \big)^{q-1}]
\label{recursion}
 \ee
The continuum limit of this recursion relation is,
 \be 
\frac{dP}{d\t}=(1-P) \ \big(  1 - (1-P)^{q-1}     \big).
\label{diffeq}
 \ee
 This equation may be integrated to give,
 \be  
P(\tau) = 1 - \lf1+ \frac{q}{N} e^{(q-1)\tau}\rg^{\frac{-1}{q-1}}
\label{master}
\ee

For early time,
\be 
P(\tau) = \frac{1}{N}e^{(q-1)\tau}.
\label{scramb}
\ee
This is the usual exponential growth of size exhibited by fast-scramblers. The scrambling time $\t_*$ is defined to be the time at which $P$ becomes of order $1.$ We see from \eqref{scramb} that,
\be 
\t_* = \frac{1}{q-1}\log{N}.
\ee

For fixed $q$ and large $N$ the early  operator growth at high temperature is given by  \cite{Roberts:2018mnp}\cite{Maldacena:2015waa},
\bea
P &\sim& e^{\omega} \cr \cr
\omega \eq \CJ t
\label{bhscram} 
\eea
In this equation $\omega$ is the dimensionless Rindler time, $t$ is ordinary \S \ time, and $\CJ$ is SYK energy scale. Comparing \eqref{scramb} with \eqref{bhscram} we obtain the relation between $\omega$ and $\t,$
\be 
\o = (q-1)\t
\label{o=(q-1)tau}
\ee

Now let us consider the late-time behavior of \eqref{master}. We find that the quantity $1-P$ exponentially decays with a quite different coefficient in the exponent,
\be 
1-P(\t) \to e^{-\t}
\label{latetime}
\ee
Later we will see that the late-time behavior is governed by the decay of quasinormal modes.

\bn

\begin{figure}[H]
\begin{center}
\includegraphics[scale=.5]{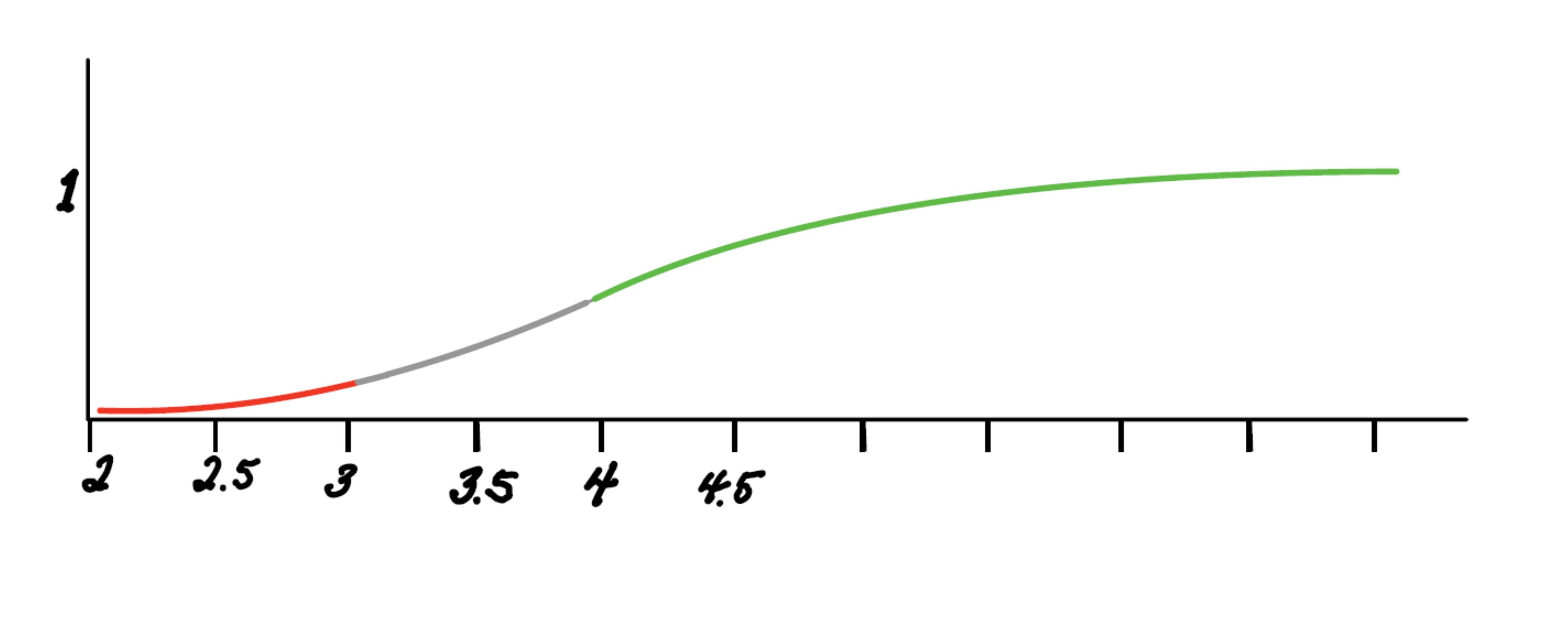}
\caption{Growth function in equation \eqref{master} for $q=4$}
\label{growth}
\end{center}
\end{figure}

Figure \ref{growth} shows a numerical plot of $P(\t)$ for the case $q=4.$
The initial early exponential growth of $P$ is shown in red in fig \ref{growth} and the final exponential approach to $P=1$ is shown in green.

\subsection{Large $q$ Limit}

\begin{figure}[H]
\begin{center}
\includegraphics[scale=.3]{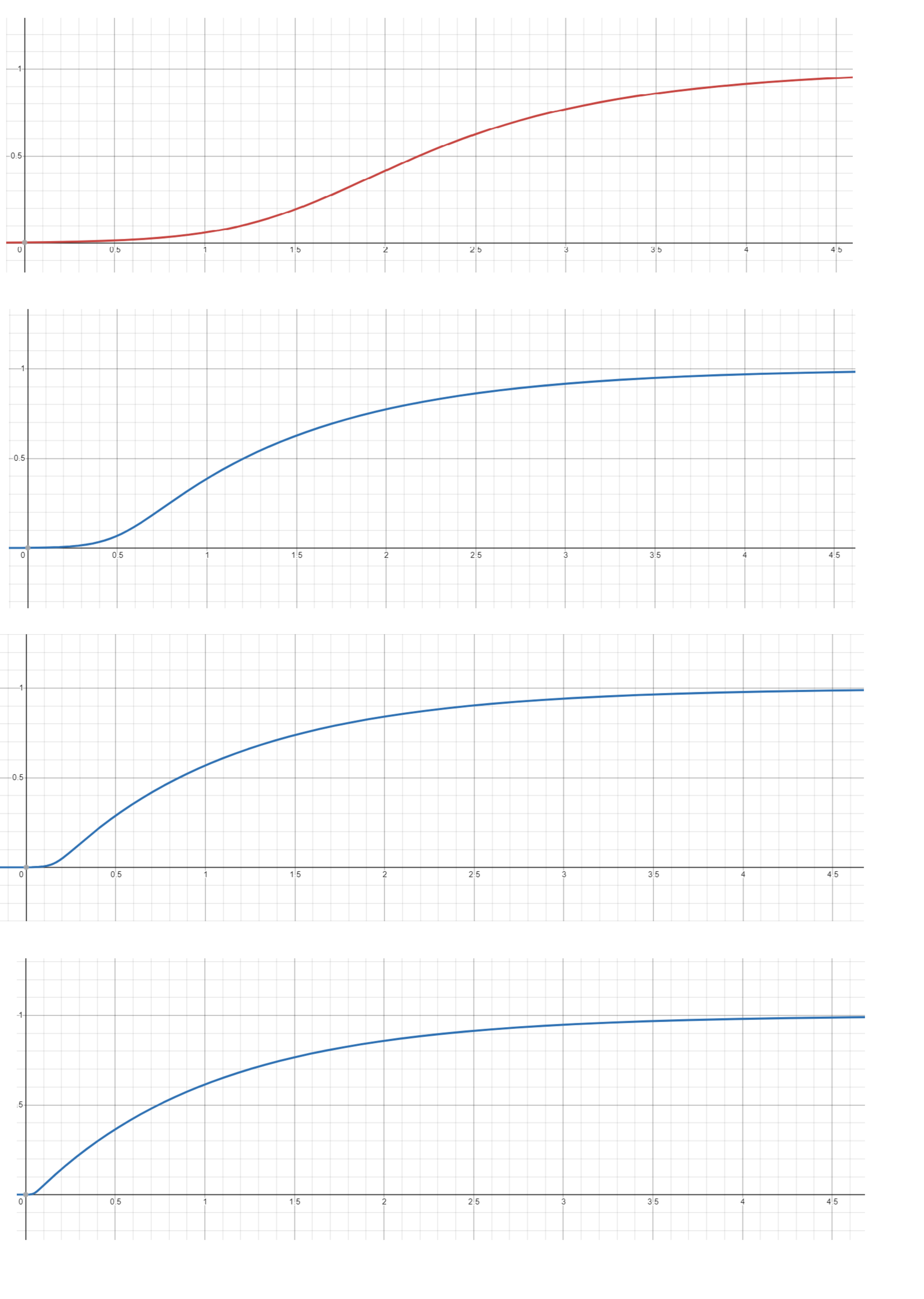}
\caption{Growth function for $q=4, 10,40,100$}
\label{growth2}
\end{center}
\end{figure}

Now let us consider the function $P$ for large $q.$ Figure \ref{growth2} shows  $P$ for $q=4$  (red curve), $q=10$ (blue), $q=40$ (green), and $q=1,000$  (purple).
Evidently the early region---the region governed by the Lyapunov behavior---shrinks to zero and the entire function  tends toward the late-time behavior \ref{latetime}. This is easy to see analytically; we leave it as an exercise for the reader.
 
In the double-scaled theory as $N$ goes to infinity so does $q.$ Thus the  function $1-P(\t)$ is a simple decaying  exponential with exponent $1.$ 
As we will see later,  this is exactly the behavior of the leading quasinormal mode.
The scrambling time is precisely the QNM  decay time,
\be 
\t_* = 1.
\label{t*=1}
\ee
There is no factor of $\log{N}.$  This behavior  is what  I called hyperfast scrambling in \cite{Susskind:2021esx}.

The bottom line is simple: the double-scaled limit of SYK is not an ordinary fast scrambler. If it has a gravitational dual then it must be something other than a black hole \footnote{Maldacena has suggested that the model may describe a black hole in a limit in which the holographic screen is placed very close to the horizon.}.

\section{QNM Decay in SYK}
Quasinormal modes of a black hole can be thought of in two ways: in terms of fields in the ``zone" outside the horizon; or in terms of deformations of  the horizon. Roughly speaking the deformations of the horizon source the fields outside the horizon. A familiar example comes from the electrodynamics of a black hole. The fields are the ordinary electromagnetic fields in the region between the horizon and the 
photon sphere, and the sources are the currents that flow on the conducting horizon. These are linearly related and both are dissipative. The QNM frequencies describe the ringing and decay of these modes. 

Earlier I claimed that in the double-scaled limit the behavior of the scrambling function in \eqref{latetime} is identical to the decay of a QNM. We can see this by  analyzing two-point functions and seeing how they decay. 

Let us consider the fermion two-point function $G_R(t)$ at infinite temperature and $t>0$.  According to \cite{Roberts:2018mnp}, in leading order of the large $q$ expansion it satisfies,
$$
G_R(t)^q  = \frac{1}{\cosh^2{\CJ t}}
$$
Here $\CJ t$ is a dimensionless measure of boundary  time that we can identify with $\omega.$ Thus we write,
\be 
G_R(\o)  =\lf \frac{1}{\cosh{\o}}\rg^{2/q}
\label{G^q}
\ee
This seems like a rather disappointing result; $G$ just goes to the time-independent value $1$ as $q \to \infty.$

But now recall \eqref{o=(q-1)tau}. Substituting $\o = (q-1)\t$ gives,
\be 
G_R(\t)  =\lf \frac{1}{\cosh{[(q-1)\t]}}\rg^{2/q}
\label{G^q2}
\ee
As $q \to \infty$ we find the non-trivial behavior,
\be 
G_R(\t) = e^{-2\t}.
\ee

This may be interpreted as the decay of a  QNM and it matches the scrambling behavior in \eqref{latetime}.

\section{Scrambling in de Sitter Space} 
In general thermalization and scrambling are different phenomena, and take place on different time-scales. Thermalization is controlled by two-point functions while scrambling is controlled by out-of-time-order four-point functions.

But 
to rephrase the result of the previous section: scrambling and thermalization are the same thing in double-scaled SYK. I now want to explain why one should think the same is true for de Sitter space.

First let's review the geometric origin \cite{Sekino:2008he}  of the logarithmic factor in the scrambling time for black holes\footnote{This section is a summary of the more technical discussion in \cite{Susskind:2018tei}\cite{Brown:2018kvn}\cite{Susskind:2019ddc}\cite{Susskind:2020gnl}}.
The holographic \dof  in AdS /CFT are located at the boundary of AdS but it is often convenient to regulate the theory so that the effective \dof are closer in. For example we can think of them as being located on a shell about one AdS length from the horizon at the edge of the near-horizon region. For a black hole in flat space a convenient location is at the photon sphere where near-horizon ``zone" gives way to the Newtonian region.

Let's suppose we perturb one of the holographic micro-\dof and ask how long it takes for the rest of them to become disturbed. The bulk mechanism is this: the perturbation creates a particle near the shell, which then falls toward the horizon. It takes a certain time to reach the stretched horizon, at which point it is quickly absorbed. The logarithmic factor $\log{S}$ or $\log{N}$ in the formula for the scrambling time is just the time that it takes for the particle to fall from the shell to the stretched horizon.

The situation for de Sitter space is different. Entropy bounds tell us that the holographic \dof should be located at the stretched horizon---not at a de Sitter distance from it \cite{Susskind:2021esx}. Therefore I expect that there will be no $\log{S}$ factor in the srambling time. 

Another way to say it is that disturbing the horizon may be described as perturbing the sources of the  QNM. The spreading of information is equivalent to the decay of those QNM, and that takes a time of order $1$ in units of the de Sitter radius. This is the hyperfast scrambling phenomenon.

To summarize: Hyperfast scrambling is not seen in the fixed-$q$ version of SYK, but it is a feature of the double-scaled limit in which $q$ grows as a power of $N$.
It is incompatible with black hole scrambling\footnote{Raphael Bousso(Private communication) has pointed out an interesting fact \cite{Emparan:2020inr}. If one considers Schwarzschild black holes in high dimensions the photon sphere moves toward the horizon as $D\to \infty.$ If one drops a particle from the photon sphere the time during which its momentum exponentially increases goes to zero. Using the size-momentum correspondence,  the Lyapunov region of exponential operator growth shrinks. This suggests  the possibility that high temperature double-scaled SYK might describe black holes in dimension $D$ which grows with the entropy. See also footnote 2.} but is expected in de Sitter space.

\end{document}